\documentclass[%reprint,
superscriptaddress,
 amsmath,amssymb,
 aps,
prb,two column
]{revtex4-1}

\usepackage{graphicx}% Include figure files
\usepackage{dcolumn}% Align table columns on decimal point
\usepackage{bm}% bold math
\usepackage{graphicx,epsf,bm,bbm,color,amsmath,ulem,amssymb,float, subfigure}

\usepackage{color}
\usepackage{ textcomp }
\usepackage{ dsfont }
\usepackage{indentfirst}
\newcommand{\remove}[1]{}

\newcommand{\bi}{\begin{itemize}}
\newcommand{\ei}{\end{itemize}}
\newcommand{\be}{\begin{enumerate}}
\newcommand{\ee}{\end{enumerate}}
\newenvironment{dfn}{{\vspace*{1ex} \noindent \bf Definition }}{\vspace*{1ex}}

	\newcommand{\beq}{\begin{eqnarray}}
	\newcommand{\eeq}{\end{eqnarray}}

\begin{document}

%\preprint{APS/123-QED}

\title{Optical response and nonlinear Hall response of twisted bilayer graphene in the insulating state}

\author{Hui Yang}
\affiliation{International Center for Quantum Materials, School of Physics, Peking University, Beijing 100871, China
}

\author{Fa Wang}
\affiliation{International Center for Quantum Materials, School of Physics, Peking University, Beijing 100871, China
}
\affiliation{Collaborative Innovation Center of Quantum Matter, Beijing 100871, China
}

\date{\today}% It is always \today, today,

\begin{abstract}
In this work, we calculate the optical response and the nonlinear Hall response of twisted bilayer graphene (TBG) in the insulating states. Different insulating states, including spin-valley polarized (SPVP) state, spin-polarized quantum Hall(SPQH) state, spin-polarized valley Hall (SPVH) state, and spin polarized Kramers-intervalley coherence (SPKIVC) state, are considered. We calculate the optical conductivity ($\sigma_{xx}$) of these four states in different experimental conditions and the linear response is different in different experimental conditions, i.e in the presence of magnetic field in different direction or in the presence of substrate. We further calculate the nonlinear Hall response which is proportional to Berry curvature dipole, and by keeping the states in hole-doped half-filling, but change the experimental conditions, the nonlinear Hall response can also help us the distinguish the insulating states. Our result can be tested in the spectroscopy and transport experiments and may be helpful to determine the nature of insulating state in TBG.
%\begin{description}
%\end{description}
\end{abstract}

\maketitle
\section{\label{sec:level1}Introduction}
Flat bands emerge in the magic angle twisted bilayer graphene (MATBG)\cite{PhysRevB.82.121407,Bis}, which can result in interesting correlation physics in this system\cite{Cao2018,cao_unconventional_2018,Xie2019,Kerelsky2019,Jiang2019,Choi2019,PhysRevLett.120.046801}. Insulating state is observed in experiment at half-filling and an superconducting state appears near this insulating state when holes or electrons are doped\cite{Cao2018,cao_unconventional_2018}. The insulating state is argued to be a Mott insulator and the superconducting state is argued to be originated from the strong correlation of electrons\cite{Cao2018,cao_unconventional_2018}. Many theories have proposed to study the origin of the superconducting and insulating states\cite{PhysRevB.99.094521,2019arXiv190308685L,PhysRevB.98.235158,PhysRevLett.121.217001,PhysRevX.8.031089,PhysRevResearch.3.013033,PhysRevB.97.235453,PhysRevX.8.041041,PhysRevB.101.224513,2020arXiv200715002L,2020arXiv200400638K} in TBG. However, the mechanism of the insulating states are unknown up to now. There are many candidate insulating states proposed for the half-filling insulating state in MATBG\cite{PhysRevX.10.031034,PhysRevLett.124.097601}. It is proposed that the insulating states can be identified by the impurity effects\cite{2020arXiv200909670G}. In this work, we propose optical and the nonlinear Hall responses to distinguish the proposed insulating state at half-filling.\\

\begin{figure}[htbp]
\centering
\includegraphics[width=15cm]{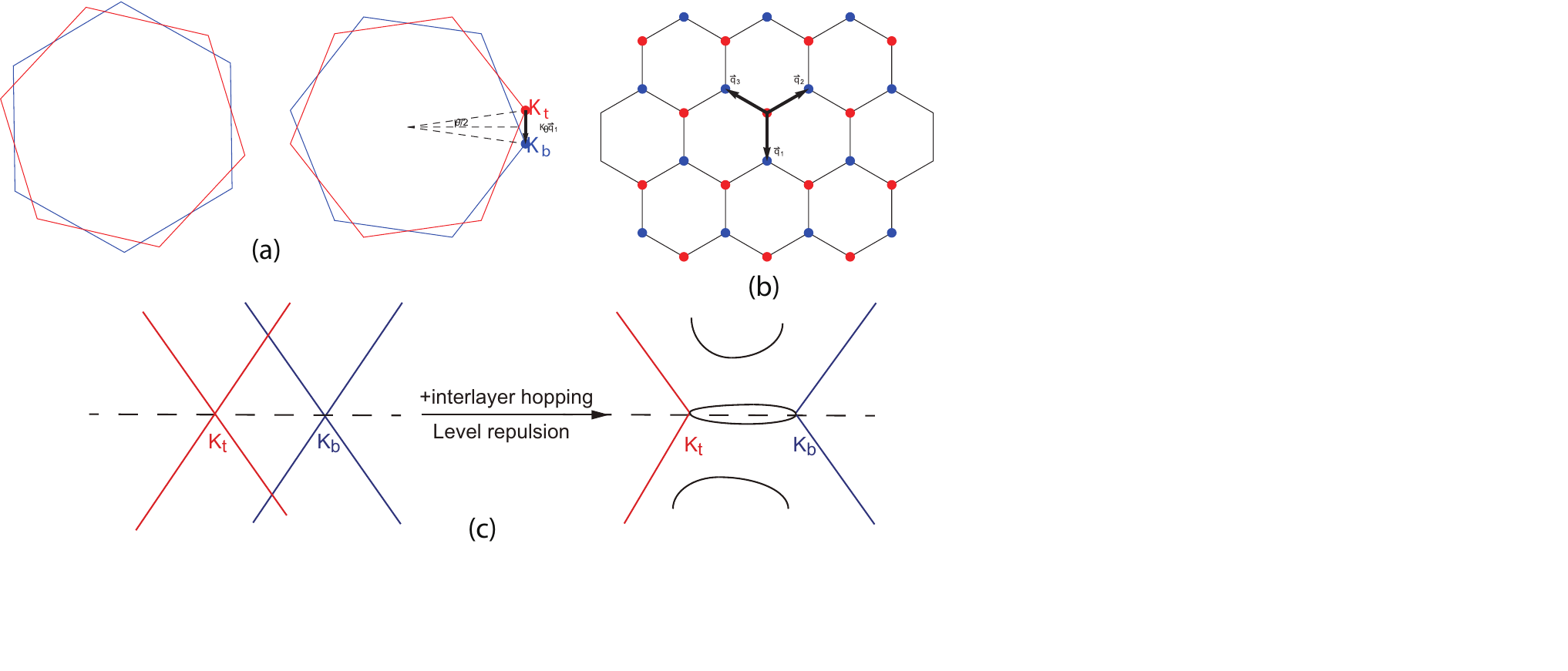}
\caption{The left (right) panel of (a) is the real (momentum) space rotation. (b) is the Mini Brillouin zone of TBG. (c) skematically describes the emergence of flat band at the magic angle.}
\end{figure}

There are many different symmetry-breaking candidate states for the MATBG at half-filling\cite{PhysRevX.10.031034}. Optical response can be an experimental method to distinguish these symmetry-breaking states, for example, a time-reversal breaking state may give rise to Hall response ($\sigma_{xy}(\omega)\neq 0$). The nonlinear Hall response (a second-order response) can also be applied to distinguish these insulating states. The nonlinear Hall effect\cite{PhysRevLett.115.216806,2020arXiv200409742D,Du2019,PhysRevLett.123.246602} originates from the anomalous velocity generated by the Berry curvature. In time-reversal invariant system, under time-reversal transformation, the Berry curvature transforms as $\Omega(k)=-\Omega(-k)$, so the total Berry curvature dipole\cite{PhysRevLett.115.216806,PhysRevLett.123.196403} is 0 after integrate over the whole Brillouin zone. However, when electric field is applied to the system, the fermi surface is shifted, and the Berry curvature is not symmetric under time-reversal, leading to a net anomalous velocity and giving rise to a Hall response.\\

In this work, we study the response in the scheme of mean-field approximation. We consider the spin-valley polarized (SPVP) state, spin-polarized quantum Hall (SPQH) state, spin-polarized valley Hall (SPVH) state, and spin polarized Kramers-intervalley coherence (SPKIVC) state. We find the behavior of conductivity and nonlinear Hall response in different experimental conditions can  help us distinguish the four insulating states. This paper is organized as follows, in sec II, we introduce the model of TBG, and add different mean field term to gap the flat band. In Sec.III, we calculate the optical conductivity and the Berry curvature dipole. We give a conclusion in Sec.IV and out main result is summarized in Table I. The details of our calculation are given in the appendix.\\

\section{\label{sec:level1}The model}
We use the Bistritzer-MacDonald model\cite{Bis} to describe low energy physics of TBG. The Hamiltonian in the $K$-valley is
\beq
H_+&=&\sum_l\sum_kf_l^\dag(k)h_k(l\theta/2)f_l(k)\nonumber\\&+&(\sum_k\sum_{i=1}^3f_t^\dag(k+q_i)T_if_b(k)+h.c),
\eeq
where $\theta=1.08^\circ$ is the magic angle, and  $f_l$ is the annihilation operator at top($l=1$) layer or bottom layer($l=-1$). $h_k(\theta)$ is the Dirac Hamiltonian and $T_i$ is the interlayer hopping\cite{Bis}, which are given in Appendix A. In our calculation, the energy $v_F|K_\theta|$ in graphene is set to unity, here $|K|$ is the magnitude of wave wave at $K_b-K_t$ given in Fig. 1(a). The $q_i$ vectors are given in Fig. 1(b). The Hamiltonian at $K'$-valley is $H_-=TH_+T^{-1}$, where $T$ is the time-reversal transformation. The total Hamiltonian of TBG is given by
\beq
H=
\begin{pmatrix}
H_+&0\\
0&H_-\\
\end{pmatrix}\otimes\sigma_0,
\eeq
	where $\sigma$ represents spin degree of freedom. The band structure is plotted in Fig. 2(b). Terms that gap the flat band of TBG Hamiltonian are given by $\Delta\sigma_z\tau_z$, $\Delta\sigma_z\eta_z\tau_z$, $\Delta\sigma_z\eta_z$, $\Delta\sigma_z\eta_y\tau_+e^{i\phi}+h.c.$ ($\tau_+=\tau_x+i\tau_y$), corresponding to the spin-valley polarized state, spin polarized quantum Hall ,spin polarized valley Hall, and spin polarized Kramers intervalley coherence\cite{PhysRevX.10.031034}, respectively, where $\sigma$, $\eta$, $\tau$ act on the spin space, sublattice space and valley space respectively. These terms is diagonal in the layer space. To consider the effects of magnetic field, we should add $B_z\sigma_z$ or $B_x\sigma_x$ for magnetic field along $z$ or $x$ direction. The term $B_z\sigma_z$ commute with the spin polarized term, while the term $B_x\sigma_x$ does not. To consider the effects of substrate, we add $V\gamma^+$, where $\gamma^+=\begin{pmatrix}1&0\\0&0\end{pmatrix}$ acts on the layer space.  In the numerical calculation, we use $\Delta=0.01$, which is the same order as the band width of the flat bands, and the qualitative result is stable with respect to $\Delta$.

\begin{figure}[htbp]
\centering
\includegraphics[width=8cm]{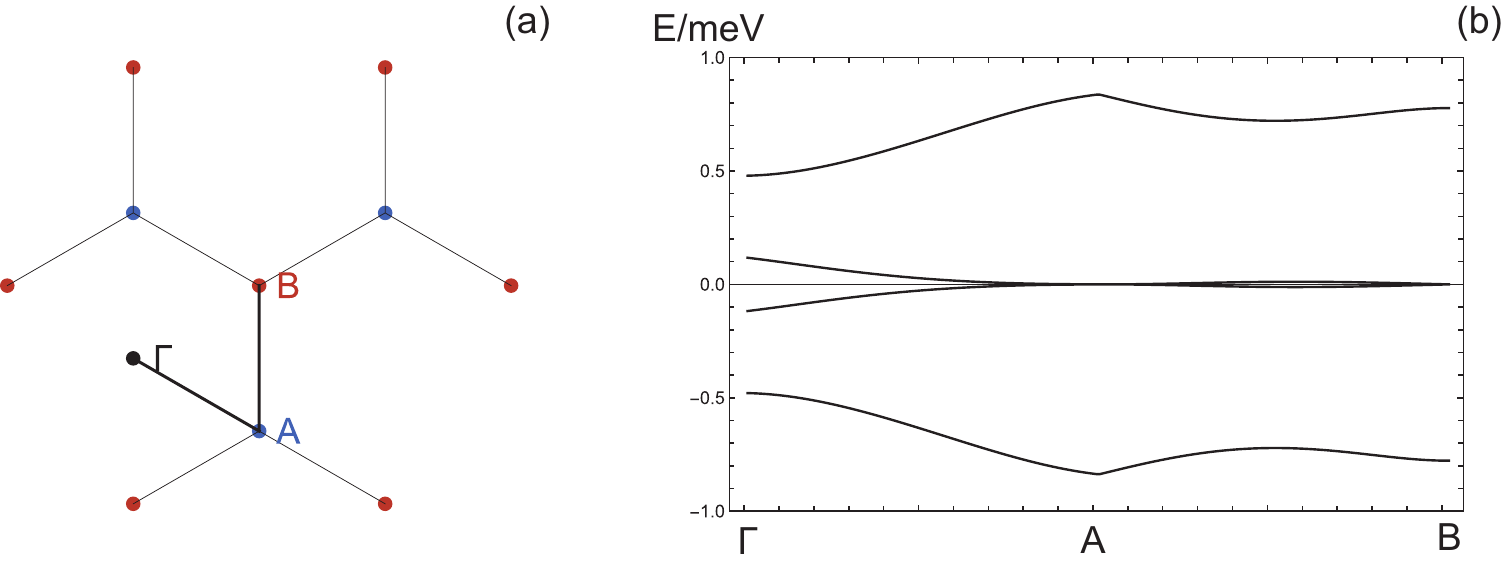}
\caption{(a) is the $10$ Dirac point in momentum space we consider in our calculation. (b) is the band structure of TBG along the line $\Gamma$A to AB in (a).}
\end{figure}

\section{\label{sec:level1}Results}
We use the Kubo formula to calculate the optical conductivity, the velocity operator $v$ is given by $\partial_{k}H$. In our numerical calculation, $e=\hbar=1$.\\

\begin{figure*}[htbp]
\centering
\includegraphics[width=16cm]{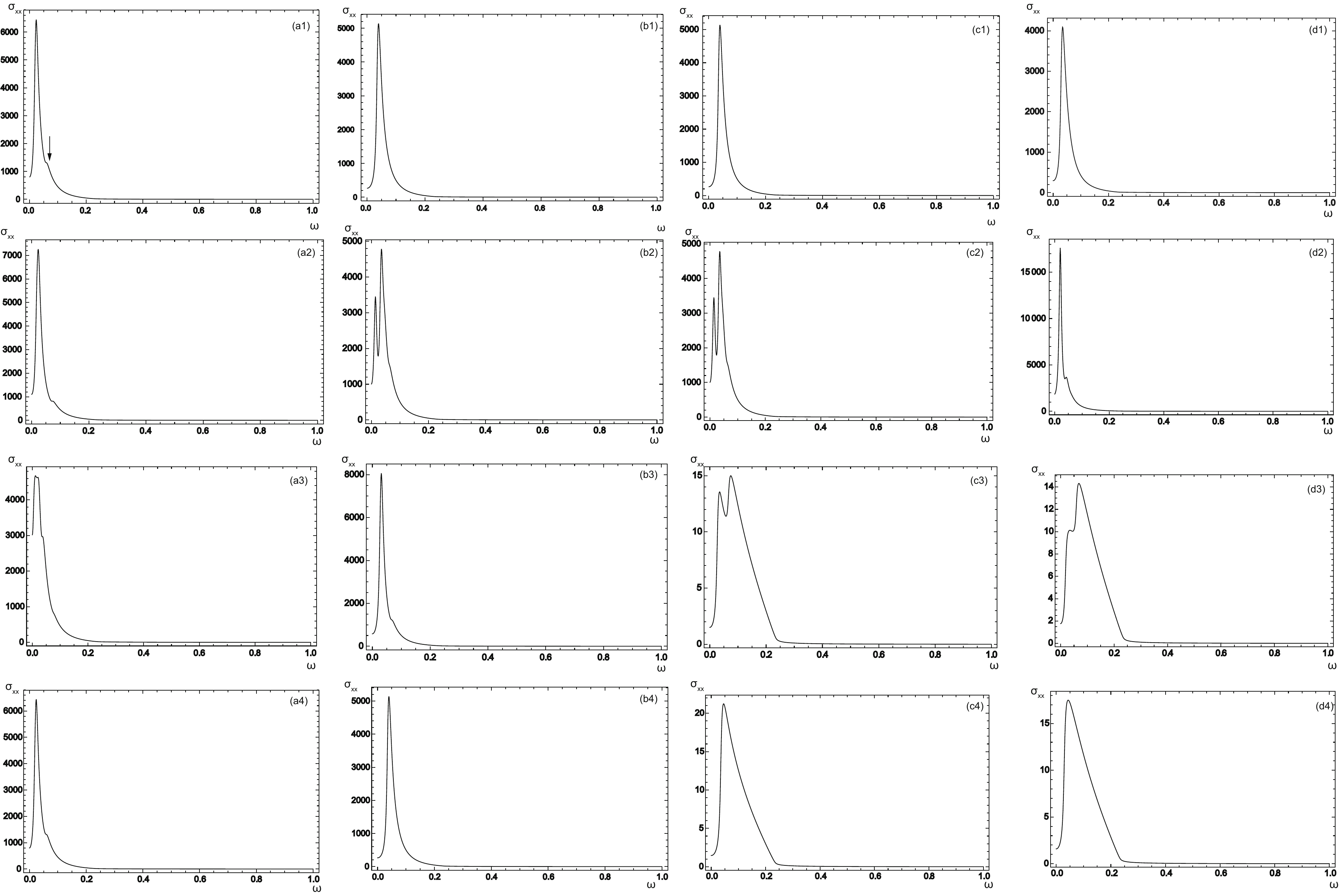}
\caption{Fig. 3 is the $\omega$ dependence of zero-temperature conductivity. (a)-(d) are the conductivities of spin valley polarized state, spin polarized quantum Hall state, spin polarized valley Hall state, and spin polarized Kramers inter-valley coherence state, respectively. (1)-(4) correspond to the optical conductivity of TBG, TBG with magnetic field $B_x$ along $x-$direction, TBG with magnetic field along $z-$direction, and TBG with substrate. From the result of $\sigma_xx$, we can see there  is a small cusp for the spin-valley polarized state. In the spin-polaried quantum Hall state, we can see two peaks in the presence of magnetic field along $x-$ direction In the spin-polarized valley and spin-polarized KIVC state, the response becomes much smaller in the presence of magnetic field $B_z$, or in the presence of substrate. In our calculation, we set $\Delta=0.01$, and in order to make the system at half filling, the chemical potential $\mu$ is plotted in Fig. 4.}
\end{figure*}

The results of optical conductivity are shown in Fig.3, from which we can distinguish the SPVP state. The two-peak feature of the $\omega$-dependence in the SPVP is different from the other three states.\\

To further distinguish the SPQH state, the SPVH state and the SPKIVC state, we calculate the nonlinear Hall response of the three states. The response function is\cite{PhysRevLett.115.216806} 
\beq
\chi_{abc}=\varepsilon_{adc}\frac{e^3\tau}{2(1+i\omega\tau)}\int_k(\partial_bf_0)\Omega_d,
\eeq
which is proportional to the Berry curvature dipole
\beq
D_{bd}=\int_k(\partial_bf_0)\Omega_d,
\eeq
here $\tau$, $f_0$ correspond to the relaxation time and zero-temperature Fermi function, respectively, and $\Omega_z$ is the Berry curvature, which reads
\beq
\Omega_z=-i(\langle\partial_xu_k|\partial_yu_k\rangle-\langle\partial_yu_k|\partial_xu_k\rangle).
\eeq
\begin{figure*}[htbp]
\centering
\includegraphics[width=16cm]{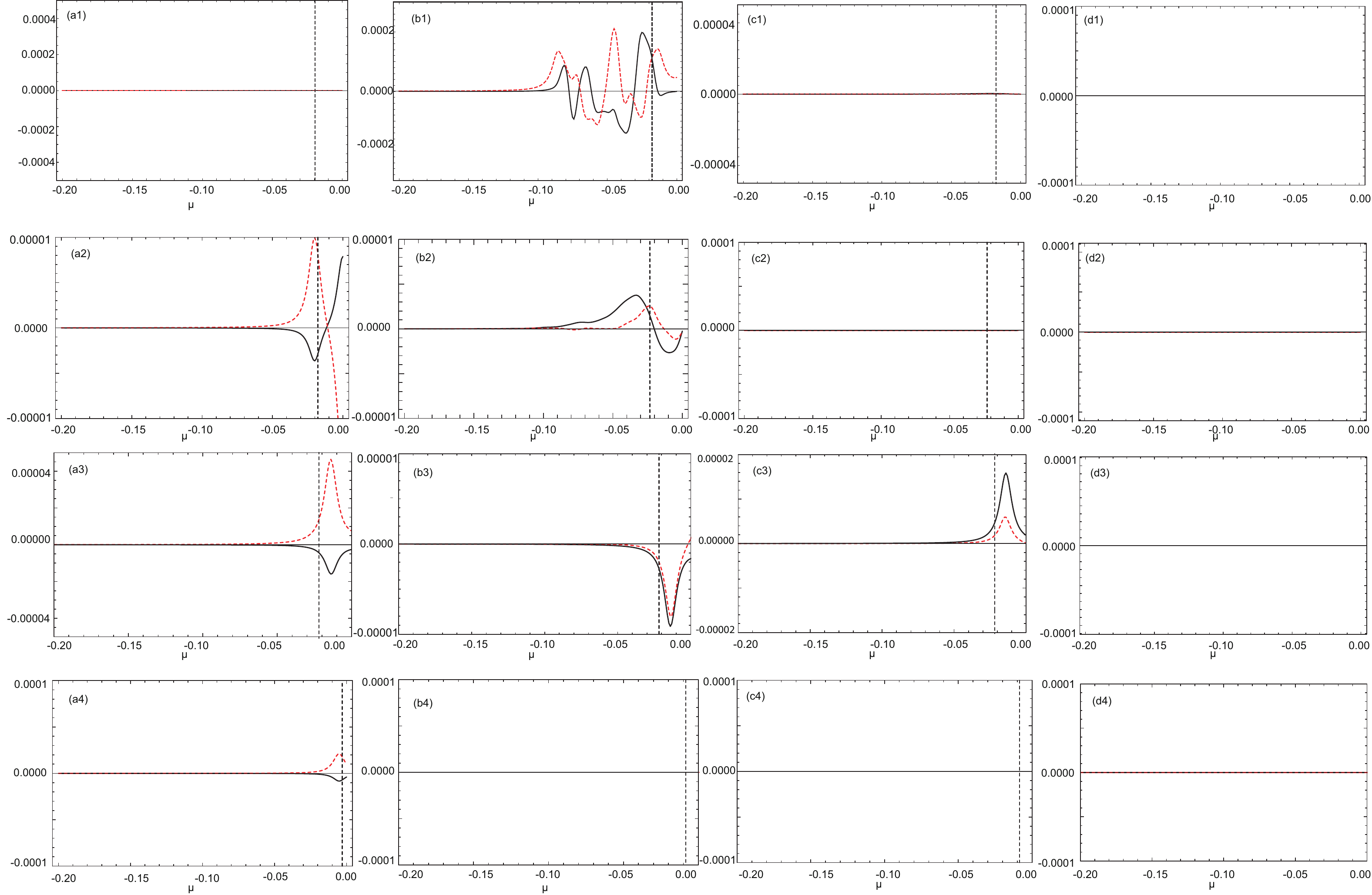}
\caption{Chemical potential dependence of Berry curvature dipole at zero temperature. The black-dashed line is the chemical potential corresponds to the hole doped half-filling (filling number $\nu=-2$).   (a)-(d) correspond to the SPVP state, SPQH state, SPVH state, and SPKIVC state, respectively. (1)-(4) correspond to the optical conductivity of TBG, TBG with magnetic field $B_x$ along $x-$direction, TBG with magnetic field along $z-$direction, and TBG with substrate.}
\end{figure*}

We ignore the relaxation time difference in different states and compare the Berry curvature dipole only. In the numerical calculation, we calculate the Berry curvature dipole $D_{bd}=\int_k(\partial_bf_0)\Omega_d$. The chemical potential dependence of Berry curvature dipole is given in Fig.4, from which we see the Berry curvature dipole is $0$ in the SPVP state and SPKIVC state. The reason of the zero Berry curvature dipole is the because of the presence of $C_2T$ or $C_2T$-like symmetry in these two states (in fact there is no time-reversal symmetry in the SPKIVC state, time reversal symmetry is breaking, but there is another anti-unitary symmetry $\tau_y\mathcal{K}$), under the $C_2T$ symmetry constraint, the Berry curvature is $0$.\\

 From the result in Fig. 4, we can distinguish the four different insulating states. In the spin-polarized KIVC state, there is always no nonlinear Hall response. The Berry curvature dipole is non-zero only in the spin-polarized quantum Hall state. While in the spin-polarized valley-polarized state, in the presence of magnetic field or substrate, the Berry curvature dipole is non-zero, and the direction of non-linear Hall current is unchanged. In the spin-polarized quantum Hall state, the nonlinear direction of Hall current is reversed between the state in the presence of magnetic field along $z-$ direction and $x-$direction. In the spin-polarized valley Hall state, the Berry curvature dipole is non-zero only in the presence of magnetic field along $z-$direction.

\section{\label{sec:level1}Conclusion}
We calculate the optical conductivity and the nonlinear Hall response to distinguish different insulating states of TBG at half-filling. The responses will be different between the state  for different experimental conditions. The optical conductivity has be detected in graphene and bilayer graphene\cite{PhysRevB.78.085432, PhysRevB.77.155409}. The nonlinear Hall effect have been observed in other materials\cite{Ma2019}. The nonlinear optical response has also been calculated in TBG\cite{PhysRevResearch.2.032015}. We believe the true state of the insulating state of TBG can be identified in future experiments.\\

\section{\label{sec:level1}Acknowledgements}

HY thanks Zhi-Qiang Gao for his helpful discussion. FW acknowledges support from The National Key Research and Development Program of China (Grant No. 2017YFA0302904), and National Natural Science Foundation of China (Grant No. 11888101).

\appendix
\section{The details of our model}
Here we review the details of our model.The Dirac Hamiltonian $h_k(l\theta/2)$ is
\beq
h_k(l\theta/2)=v_F\begin{pmatrix}
0&(k_x+ik_y)e^{il\theta/2}\\
(k_x-ik_y)e^{-il\theta/2}&0
\end{pmatrix}.
\eeq
The hopping matrices are given by
\beq
T_1&=&\begin{pmatrix}
0&1\\
1&0
\end{pmatrix},\\
T_2&=&
\begin{pmatrix}
0&e^{-2\pi i/3}\\
e^{2\pi i/3}&0
\end{pmatrix},\\
T_3&=&
\begin{pmatrix}
0&e^{2\pi i/3}\\
e^{-2\pi i/3}&0
\end{pmatrix}.
\eeq

 As shown in Fig. 2(a), we consider hoppings between these $10$ $K(K')$ points. Considering all the degeneracies there are $8$ flat bands. The projection operator of the flat band is $|\psi\rangle$, with
\beq
|\psi\rangle=
\begin{pmatrix}
|u_1(k)\rangle&|u_2(k)\rangle&\cdots&|u_8(k)\rangle
\end{pmatrix}^{T},
\eeq
where $|u_i(k)\rangle$ is the Bloch wave function of the flat band.

In our calculation, we first project the total Hamiltonian to the flat-band subspace, and then calculate the Bloch wave functions and energies. Finally, we can calculate the optical conductivity and the Berry curvature dipole using the Kubo formula and Eq. (3).

\bibliographystyle{apsrev4-1}

\bibliography{OC.bib}
\end{document}